\documentclass[preprint,aps]{revtex4-1}

\usepackage{color}
\usepackage{mathtools}
\usepackage{array}
\usepackage{tabularx}
\usepackage{diagbox}
\usepackage{tikz}
\usetikzlibrary{decorations.pathmorphing}
\usepackage{bm}
\usepackage{amsmath,bm,amssymb,amsfonts,dcolumn,color,graphicx,latexsym,epsfig}
\usepackage{booktabs} 
\usepackage{subfig}
\usepackage{graphicx, slashed}
\usepackage{float}

\usepackage[absolute]{textpos}

\usepackage{tikz}
\usepackage{amsmath}

\usepackage{multirow}
\usepackage{natbib}

\usepackage{float}

\usepackage{booktabs}
\usepackage{pifont}
\usepackage{mathrsfs}
\usepackage{caption}
\usepackage{caption, threeparttable}
\begin{document}

\title{\Large \bf Scalar and Spinor Quasi Normal Modes of a 2D Dilatonic Blackhole}

\author{Pabitra Gayen} \email{pabitra.rs@presiuniv.ac.in} \author{Ratna Koley} \email{ratna.physics@presiuniv.ac.in}
\affiliation{Department  of  Physics,  Presidency  University,  86/1 College Street,  Kolkata  700073,  India.}


\begin{abstract}{External non-minimally coupled scalar and spinor field perturbations have been studied in a (1 + 1) 
dimensional dilatonic blackhole spacetime \cite{Mandal:1991tz, Witten:1991yr}. Exact analytical expressions of the quasi-normal mode 
frequencies have been found for both the cases. In the scalar perturbations the quasi-normal mode frequencies turn out to be purely 
imaginary and negative. Furthermore 
we have found that the quasi-normal frequencies for Dirac field exhibit both real and imaginary part. The QNM frequencies decay 
monotonically with the overtone number under certain class of the blackhole parameters. The decay profile ensures the stability of the 
blackhole spacetime under these perturbations.}\end{abstract}


\maketitle

\section{Introduction}\label{sec1}

Isolated black holes are just theoretical
artifacts. In general, the blackholes are likely to be surrounded by perturbing 
scalar, vector, spinor and similar fields. The perturbed blackhole spacetimes are intrinsically dissipative 
due to the presence of the event horizon and the eigen modes of these systems are known as quasi-normal modes (QNMs). These 
modes possess a set of well defined complex frequencies known as quasi-normal frequencies (QNFs) \cite{Kokkotas:1999bd}. 
We are particularly interested in studying the QNMs as they provide the window for an asymptotic observer 
to know the microscopic behaviour of blackholes (BHs).   In recent times the QNMs are being widely studied 
in the context of gravitational wave astronomy as they are capable of giving precise information about the 
parameters of the blackhole as well as they can serve as a new tool to test the theories of gravity \cite{LISA:2022kgy}. 
The QNM research has been expanded to encompass diverse topics like analogue gravity, alternative theories of gravity, 
higher dimensional spacetime etc \cite{Chakraborty:2017qve,Kanti:2005xa,Birmingham:2001pj,Kovtun:2005ev,Seahra:2004fg}. 
Recent studies \cite{Verlinde:2016toy,Tuveri:2019zor,Dvali:2020wqi,Dvali:2013eja,Cadoni:2020mgb} 
also suggest that some of the major outstanding puzzles of modern cosmology - dark energy, dark matter or modified 
Newtonian dynamics at galactic scales, information loss beyond the blackhole horizon can be addressed by assuming 
the underlying microscopic quantum theory of gravity executing a multi-scale behaviour \cite{Cadoni:2021qfn}. 

The work of Hawking and Penrose in the 1960s \cite{Hawking:1973uf} showed the classical theory of gravity, {\sc gr}, breaks down at certain regimes like the singularities 
at the cosmological big bang or gravitational collapse into black holes. 
The presence of singularities and event horizons, also leads to paradoxical situations such as information loss \cite{Preskill:1992tc}. It is hoped that such paradoxes of the classical theory can be cured by quantizing gravity, which needs introduction of higher dimension. Such theories met with moderate success. 
On the other hand, {\sc gr} being a non-linear theory and because of its inherent complexity, in four dimensions the exact analytical solutions are not easy to obtain. Classical {\sc gr} becomes very much simpler 
in spacetime dimensions less than four, and this simplification carries over to the quantum theory as
well. Although the lower dimensional models are not physical in the 
true sense of the term,  the mathematical techniques and general reasoning 
are much the same as those for the full four-dimensional theory. In many situations, the
understanding of a lower dimensional theory does provide
important insights into the features of the actual four 
dimensional theory -- for example, they have been proved to be very useful for investigating the blackhole thermodynamics
\cite{Kettner:2004aw, Bhattacharjee:2020nul, Grumiller:2002nm, MANN1990134}. Many problems associated to 
full four dimensional theory can be addressed with much more control in lower dimensional toy models.

In this article we choose to work with a 2D line element that arises in the context of a stringy black hole proposed by 
Mandal-Sengupta-Wadia (MSW) \cite{Mandal:1991tz} and also by Witten \cite{Witten:1991yr}.  Without going into the details of how this line element is obtained 
we explore some intriguing features treating the geometry is given to us. However, interested reader may refer to the 
original references \cite{Mandal:1991tz, Witten:1991yr} for necessary rigor. The geodesics and geodesic deviations in this geometry 
have been extensively studied in \cite{Koley:2003tp}. 
Several other aspects of classical and quantum gravity in MSW black hole spacetime in $(2+1)$ dimensions have been 
explored in \cite{Fernando:2004ay,Fernando:2003ai,Chan:1994qa,Sebastian:2014dka,Lopez-Ortega:2011ysu} which include 
QNMs for minimally coupled massless and massive scalar field, Dirac field by using WKB approximation. In summary, 
the two dimensional blackhole spacetimes work as a great pedagogical tool. The two dimensional JT dilatonic black 
holes are quite well explored in the context of QNMs of  non-minimally coupled perturbing scalar fields 
\cite{Zelnikov:2008rg,Cadoni:2021qfn,Bhattacharjee:2020nul}. 
In the present work we picked up MSW \cite{Mandal:1991tz} and tried to understand the basic features of the 
perturbations of a single horizon blackhole in 2D dilatonic gravity corresponding to the scalar and spin half fields. 
We studied perturbations due to {\it{non-minimally coupled}} massless and massive scalar fields with the dilatonic field. 
It is important to note that the coupling parameter plays an important role in the nature of QNFs and the stability of the 
blackhole geometry. It is crucial to note that  exact analytical expressions for the QNM 
frequencies have been achieved in all the cases. Non-minimally coupled massless spin-half 
perturbations leads to QNFs which have both real and imaginary parts.

\section{Quasi-normal modes for massive scalar perturbations}\label{sec2}
It is well known 
that Einstein's gravity is purely topological in two spacetime dimensions since the Einstein-Hilbert
action is just the Gauss-Bonnet topological term. The presence of further structures like the dilaton field can invoke  
the dynamics. We choose to work with the first stringy black hole discovered in the
early 1990s by Mandal et al. \cite{Mandal:1991tz} and simultaneously by Witten \cite{Witten:1991yr}  which arises in a low-energy 
effective description of dilatonic black holes in string theory:

\begin{equation}\label{action}
S = \int d^2x \sqrt{-g} e^{-2\phi}\left [ R + 4\left (\nabla \phi\right )^2
+4\Lambda^2 \right ] 
\end{equation}

where $\phi$ is the dilaton field and $\Lambda^2$ is the cosmological constant.
The spacetime is described by the following line element in the geometric unit with $G = 1$ and $c = 1$.

\begin{equation}\label{line element}
ds^2=\left( 1-\frac{M}{r} \right) dt^2-\frac{kdr^2}{4r^2 \left( 1-\frac{M}{r} \right)} 
\end{equation}

 For the completeness, let us mention that $M$ is connected to 
the mass of the black hole, $k$ to the central charge parameter and $\phi$ to the dilaton field 
(with $e^\phi=\sqrt{\frac{M}{r}}$). 
As apparent from the metric, the horizon is at $r = M$ as $g_{tt} \rightarrow 0$ and $g_{rr} \rightarrow \infty $. 
The domain of our concern lies in the range $r > M$ in radial direction and $-\infty < t < \infty$ for time coordinate. 
Note that the spacetime is very similar to the $t-r$ section of the static, spherically symmetric Schwarzschild metric. 
The horizon is a point unlike a spherical surface in the case of a four dimensional blackhole. It is important to note that the 
above line element can be reduced to the familiar 2D dilatonic BH metric with a coordinate transformation $\bar r = \frac{\sqrt k}{2}\ln r$

\begin{equation}{\label{2d:lelem}}
ds^2 = -g(\bar r)dt^2 + \frac{d{\bar r}^2}{g(\bar r)}
\end{equation}

where $g(\bar r) = 1-M \exp(-2{\Lambda {\bar r}})$. Let us now give some other 
examples of black holes in two dimensional dilaton gravity. In the sense of full String Theory 
we note the exact metric  

\begin{equation}{\label{2d:verlinde}}
ds^{2} = 2 (k - 2) [-\beta(r) dt^2 + dr^2]
\end{equation}

where $\beta(r) = (\coth^2 r -2/k)^{-1}$ as the dilaton is given as 
$\phi = \phi_{0} + \frac{1}{2} \mbox{ln} \vert \sinh^2 \frac{2r}{\beta} \vert$.
For $k \rightarrow \infty$ limit the 
above metric reduces to the one obtained by Witten in ~\cite{Witten:1991yr}:

\begin{equation}{\label{2d:wit}}
ds^{2} = -\tanh^2 r~dt^2 + dr^2
\end{equation} 

 This can also be obtained from the metric by Mandal {\em et.al.}
\cite{Mandal:1991tz} by a 
simple coordinate transformation. In this paper we choose to work with the line element given in Eq. (\ref{line element}).

Let us now consider an external  scalar field $\Phi$ as the perturbing field. 
We have considered a generic coupling factor $h(\phi)$ between the dilaton and the perturbing field to ensure that 
the dynamics is properly captured in the Klein-Gordon equation \cite{Kettner:2004aw}. 
For a massive scalar perturbation the Klein-Gordon equation in a general 
curved background is given by 

\begin{equation}\label{eq: KG curved spacetime}
\frac{1}{\sqrt{-g}h(\phi)}\partial_\mu\left(\sqrt{-g}h(\phi)g^{\mu \nu}\partial_\nu \Phi\right)+m^2\Phi=0
\end{equation}
${}$ \\

Since the metric coefficients - sourced by the background dilaton field - are function of $r$
only, we can choose $h(\phi) = h(r)$ without loss of generality. 
Thus the time dependent part of the above equation will exhibit a harmonic nature. We now decompose $\Phi (r,t)$ in the following form 

\begin{equation}\label{MS modes}
\Phi(r,t)=\frac{R(r)}{\sqrt{h(r)}}\,e^{-i\omega t}
\end{equation}

to reduce the Eq. (\ref{eq: KG curved spacetime}) in a form effectively analogous to a parametric 
harmonic oscillator, where $\omega$ is the frequency. The spatial wave function, $R(r)$, is defined 
in such a way that we achieve the simplest possible wave equation. Following the standard methodology let us introduce 
a generalized tortoise coordinate for the spacetime given in Eq. (\ref{line element}) as 

\begin{equation}\label{tortoise coordinate}
r_*=\int \frac{dr}{\frac{2(r-M)}{\sqrt{k}}}=\frac{\sqrt{k}}{2}\ln \left( \frac{r-M}{\sqrt{k}} \right)
\end{equation}

Note that, this transformation maps the horizon and the spatial infinity of radial coordinate 
into $r_*\rightarrow -\infty$ and $r_*\rightarrow \infty$ respectively. 
Under this coordinate transformation the radial part of Eq. (\ref{eq: KG curved spacetime}) reduces to an equation analogous to the 
Schrodinger equation 

\begin{equation}\label{eq: tortoise}
\frac{d^2R}{dr_*^2}+\left[{\omega}^2-V_{eff}(r_*)\right]R=0
\end{equation}

In radial coordinates the effective potential has the following form 

\begin{equation}\label{pot: radial}
V_{eff}(r)=\frac{2(r-M)}{kh(r)}\left( \left( r-M \right) h''(r) + h'(r) - 
\frac{(r-M)}{2h}{\left( h'(r) \right)}^2 \right)+\left( 1-\frac{M}{r}\right)m^2   
\end{equation}

where prime denotes derivative w.r.t. $r$. 
The coupling term, $h(r)$ plays an important role in generating the modes of oscillation. It is apparent from 
the above expression that the massless modes are not supported without the coupling term. In fact, for a constant 
coupling the potential vanishes for massless case. 
A judicious choice of the function $h(r)$ leads to interesting features in the QNM spectrum. 
Implementing a power law behaviour, $h(r)\propto r^\sigma $, the potential in Eq. (\ref{pot: radial}) reduces to the following 
form in tortoise coordinates

\begin{equation}\label{pot: tortoise}
V_{eff}(r_*)=\frac{\sigma(\sigma-2)}{k}{\xi}^2+\left( \frac{2\sigma}{k}+m^2 \right) {\xi}
\end{equation}

where

\begin{equation}\label{xi}
\xi \equiv \frac{\sqrt{k}e^{2r_*/\sqrt{k}}}{M+\sqrt{k}e^{2r_*/\sqrt{k}}}.
\end{equation}

\begin{figure}[ht]
 \centering
   \includegraphics[width=6cm, height=4 cm]{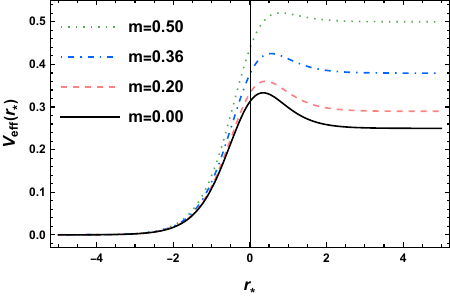}
  \caption{Effective potential experienced by a massive scalar field in 
  tortoise coordinates. }
 \label{fig:V_E}
\end{figure}

The effective potential has been plotted in Figure \ref{fig:V_E} for the choice of parameters 
$\sigma=\frac{1}{2}$, $M=1$, and $\sqrt{k}=1$. In this figure we have shown the effective potentials 
experienced by fields having different masses. The figure depicts that effective potentials go to zero at 
the horizon and take different asymptotic values (depending on the mass of the fields) at spatial infinity.
The nature of the potential may vary with the relative strengths of the blackhole parameters 
and the non-minimal coupling.  

We now find the exact expressions for the QNM spectrum. They are characterised by the complex frequencies, 
the imaginary part of which causes damping of the modes.
In general this can be obtained either by solving the Eq. (\ref{eq: tortoise}) requiring the Dirichlet boundary conditions at 
infinity or by considering the modes are purely ingoing at the horizon \cite{1985ApJ...291L..33S}. At this point, we impose
yet another transformation of coordinates to include the whole range ($-\infty<r_*<\infty$) of tortoise coordinate 
\begin{equation}\label{x coordinate}
x=\frac{1}{2}\left( 1+\tanh{\frac{r_*}{\sqrt{k}}} \right)
\end{equation}

This change maps horizon to $x=0$ and spatial infinity to $x=1$.
The functional form of the effective potential (\ref{pot: tortoise}) in the new coordinates is given by 

\begin{equation}\label{pot: x coordinate}
V_{eff}(x)=\frac{\sigma(\sigma-2)x^2}{{\left(M(1-x)+\sqrt{k}x\right)}^2}+
\left( \frac{2\sigma}{k}+m^2\right){\left(\frac{\sqrt{k}x}{M(1-x)+\sqrt{k}x}\right)}
\end{equation}

We rewrite the Eq. (\ref{eq: tortoise}) in the new coordinates to have the following form
\begin{equation}\label{eq: in x}
\frac{4x^2(1-x)^2}{k}\frac{d^2R}{dx^2}+\frac{4x(1-x)(1-2x)}{k}\frac{dR}{dx}+\left({\omega}^2-V_{eff}(x)\right)R=0
\end{equation}

According to the choice of the boundary condition, $R(x)$ is purely ingoing at the horizon and 
outgoing at spatial infinity. Let us construct $R(x)$ in the following way to achieve exact solutions of the above equation. 

\begin{equation}\label{R in x coordinate}
R(x)=x^\alpha{(1-x)}^\beta F(x)
\end{equation}

with $\alpha$ and $\beta$ satisfying 

\begin{equation}\label{cond: hypergeometric I}
4{\alpha}^2+k{\omega}^2=0 \qquad \qquad \qquad km^2+\sigma^2-4\beta^2-k\omega^2=0
\end{equation}

where we have imposed a constraint $M=\sqrt{k}$ on the blackhole model parameters 
to achieve exact analytical results. The Eq. (\ref{eq: in x}) reduces to a standard hypergeometric equation \cite{abramowitz1964handbook,wang1989special} for $F(x)$ with the above construction (Eq. (\ref{R in x coordinate}))

\begin{equation}\label{eq: Hypergeometric DE for MSF}
x(1-x)\frac{d^2F}{dx^2}+\left( \tilde{c}-\left( \tilde{a}+\tilde{b}+1 \right) x \right) \frac{dF}{dx}-\tilde{a}\tilde{b} F=0
\end{equation}

where
\begin{equation}
\tilde{a}=\alpha+\beta+1-\frac{\sigma}{2} \quad \quad \quad \tilde{b}=\alpha+\beta+\frac{\sigma}{2} \quad \quad \quad \tilde{c}=2\alpha+1.
\end{equation}

We consider the following relations for the parameters $\alpha$ and $\beta$, to study $\Phi$ in detail 

\begin{equation}
\alpha=i\frac{\omega \sqrt{k}}{2} \qquad \qquad \qquad \beta=\frac{1}{2}\sqrt{km^2+\sigma^2-k\omega^2}
\end{equation}

The solution of Eq. (\ref{eq: Hypergeometric DE for MSF}) is given by the standard hypergeometric function of second kind in Eq. (\ref{sol: hypergeometric for SF }) for $\tilde{c}$ not being  an integer
\begin{equation} \label{sol: hypergeometric for SF }
F(x) = C_1 \, {}_{2} F_{1}(\tilde{a},\tilde{b};\tilde{c};x) + C_2 \, x ^{1-\tilde{c}}  {}_{2} F_{1}(\tilde{a}-\tilde{c}+1,\tilde{b}-\tilde{c}+1;2-\tilde{c};x)
\end{equation}
where $C_1$, $C_2$ are constants and ${}_{2}F_{1}$ is the hypergeometric function. We now impose the boundary conditions (i) purely ingoing near the horizon,  (ii) purely outgoing for $x \rightarrow 1$ to calculate the QNMs.
Purely ingoing behaviour of the solution near the horizon reduces $R(x)$ in the following form

\begin{equation}
R(x)=C_2~x^{-i\omega \sqrt{k}/2}(1-x)^{\sqrt{km^2+\sigma^2 -k\omega^2}/2}~{}_{2}F_{1} (\tilde{a}-\tilde{c}+1,\tilde{b}-\tilde{c}+1;2-\tilde{c};x)
\end{equation}

where we have set $C_1=0$ for the solution to be compatible with the boundary condition.
The scalar quasinormal modes are purely outgoing at spatial infinity. We now  use the Kummer's property \cite{abramowitz1964handbook,wang1989special} of the hypergeometric function for further calculation

\begin{equation}\label{kummer's property}
\begin{split}
{}_2F_1(\tilde{a},\tilde{b};\tilde{c};x) &= \frac{\Gamma(\tilde{c}) \Gamma(\tilde{c}-\tilde{a}-\tilde{b})}{\Gamma(\tilde{c}-\tilde{a}) \Gamma(\tilde{c} - \tilde{b})} {}_2 F_1 (\tilde{a},\tilde{b};\tilde{a}+\tilde{b}+1-\tilde{c};1-x) \\
&+ \frac{\Gamma(\tilde{c}) \Gamma( \tilde{a} + \tilde{b} - \tilde{c})}{\Gamma(\tilde{a}) \Gamma(\tilde{b})} (1-x)^{\tilde{c}-\tilde{a} -\tilde{b}} {}_2F_1(\tilde{c}-\tilde{a}, \tilde{c}-\tilde{b}; \tilde{c} + 1 -\tilde{a}-\tilde{b}; 1 -x)
\end{split}
\end{equation}

for $\tilde{c}$ is not a negative integer \cite{abramowitz1964handbook} and $\tilde{c}-\tilde{a}-\tilde{b}$ is not an integer. In the limiting case when $r_* \to + \infty$ ($x \to 1$), imposing the above property of the hypergeometric function, $R(x)$ is reduced to

\begin{equation} \label{eq: radial infinity}
R \approx  \frac{\Gamma(\tilde{C})\Gamma(\tilde{C}-\tilde{A}-\tilde{B})}{\Gamma(\tilde{C}-\tilde{A})\Gamma(\tilde{C}-\tilde{B})} e^{-\sqrt{km^{2}+\sigma^{2}-k\omega^{2}} r_*} + \frac{\Gamma(\tilde{C})\Gamma(\tilde{A}+\tilde{B}-\tilde{C})}{\Gamma(\tilde{A})\Gamma(\tilde{B})}e^{\sqrt{km^{2}+\sigma^{2}-k\omega^{2}} r_*} 
\end{equation} 

where
$
\tilde{A} = \tilde{a}-\tilde{c}+1, 
\tilde{B} = \tilde{b}-\tilde{c}+1, 
\tilde{C} = 2-\tilde{c}.$ 
Applying the boundary condition that $F(x)$ is purely outgoing at the asymptotic limit is 
therefor equivalent to imposing the condition
\begin{equation}
\tilde{C}-\tilde{A}= -n \qquad \text{or} \qquad \tilde{C}-\tilde{B} = -n,
\end{equation}

where $n  = 0, 1, 2, \dots $ is the  overtone number. We thus get the quasinormal frequencies (QNFs) for the massive scalar field $\Phi$ as 

\begin{equation} \label{QNFs massive scalar field}
\omega^{\prime} = -i\frac{\left( 4n^2+4n\sigma -km^2 \right)}{2\sqrt{k}\left( 2n+\sigma \right)} \qquad \qquad
\omega^{\prime \prime} = -i\frac{\left( 4\left( n+1 \right) \left( n+1-\sigma \right) -km^2 \right)}{ 2\sqrt{k} \left( 2n+2-\sigma  \right)}
\end{equation}

\begin{figure}[ht]
 \centering
   \includegraphics[width=8.5 cm, height =4 cm]{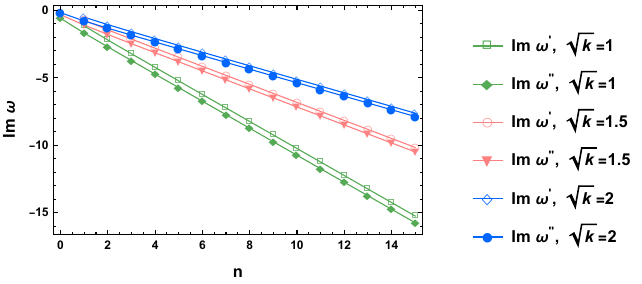}
  \caption{Variation of imaginary part of QNFs for massive ($m=0.5$) scalar field with $n$ for different  $\sqrt{k}$ values. Considering the values of the parameters $M=1$ and $\sigma=0.5$.}
 \label{fig:QNFsMSF}
\end{figure}

We have shown the variation of the QNFs with overtone number in Figure \ref{fig:QNFsMSF} where we have considered the imaginary part of the QNF is negative. As is well known,  for the stability of the modes in Eq. (\ref{MS modes}) the amplitude should decay with time at any fixed spatial position. The modes will lead to stability under the condition 
\begin{equation}
n>\frac{1}{2}\left(\sqrt{km^2+\sigma^2}-\sigma \right) \quad \text{for} \quad n \ge 1  
\end{equation}

It is worth noting that the frequency $\omega'$ will not give any stable mode for $n = 0$ whereas one can achieve stability for $\omega''$ under the condition $2>\sqrt{km^2+\sigma^2}+\sigma $.

\begin{figure}[ht]
\centering
\includegraphics[width=8.5 cm, height =4 cm]{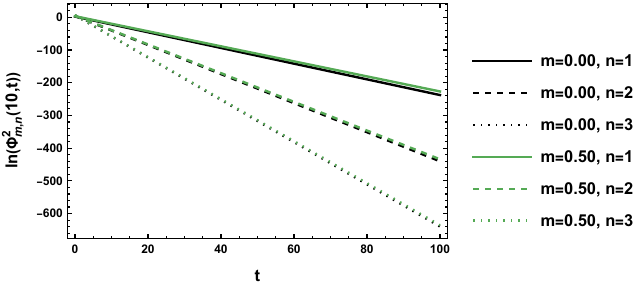}
\caption{Time evolution of square  of different massive scalar modes in logarithmic scale at $r=10$ corresponding to the frequency $\omega '$. Considering the value of the parameter $\sqrt{k}=1$.}
\label{fig: scalar mode variation wI}
\end{figure}

\begin{figure}[ht]
\centering
\includegraphics[width=8.5 cm, height =4 cm]{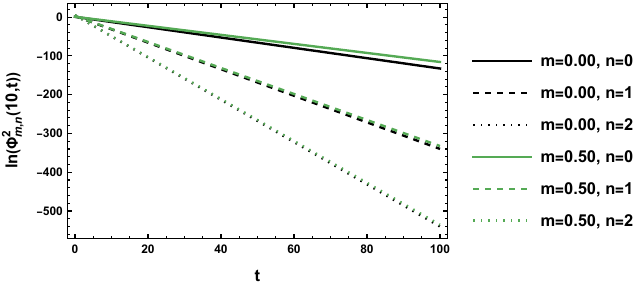}
\caption{Time evolution of square  of different massive scalar modes in logarithmic scale at $r=10$ corresponding to the frequency $\omega ''$.}
\label{fig: scalar mode variation wII}
\end{figure}

Time evolution of square  of different massive scalar modes in logarithmic scale at $r=10$ is shown in the Figure \ref{fig: scalar mode variation wI} and \ref{fig: scalar mode variation wII}. It is seen that as the mass of the field increases the mode becomes less and less damped for a fixed value of $n$. It is also clear that the damping of a particular massive mode increases with the increase of $n$.  


\section{Quasi-normal modes for massless Dirac field non-minimally coupled with dilaton field}
We propose a scalar-fermion Yukawa type, non-minimal coupling between the dilaton and the Dirac field in the background geometry of Eq. (\ref{line element}). The action for a massive Dirac field with this coupling is given by

\begin{equation}
S=\int d^2x \sqrt{-g}  \left[ \Bar{\Psi} \left( i \gamma^\mu \nabla_\mu - m \right) \Psi  -  \zeta (r) \phi \Bar{\Psi} \Psi \right ]
\end{equation}

The term, $\zeta (r)\phi$ can be written as $m_{eff}$ in the following action in Eq. (\ref{action}) by considering $m_{eff} =\zeta (r)\phi$. For further exploration we have chosen a power law behavior for $m_{eff}$ as $Qr^\sigma$ because the dilaton field is function of radial coordinate only. Thus we see from the above action that a massless Dirac field also picks up an effective mass parameter due to the interaction term and the above action reduces to the following form

\begin{equation}\label{action}
S=\int d^2x \sqrt{-g} \Bar{\Psi} \left( i \gamma^\mu \nabla_\mu - m_{eff} \right)  \Psi
\end{equation}

Variation of the above action with respect to $\Bar{\Psi}$ leads to the equation of motion of the Dirac field as
\begin{equation}\label{dirac equation cst}
i \slashed{\nabla}\Psi=m_{eff}\Psi
\end{equation}

where $\Psi$ is a two-spinor, $\slashed{\nabla}$ $(=\gamma^\mu\nabla_\mu)$ is Dirac slash operator with 
$\nabla_\mu$ $(=\partial_\mu-\Gamma_\mu)$  being the covariant derivative and $\Gamma_\mu$ is the spin connection. We write the metric in Eq. (\ref{line element}) in tortoise coordinate as given in Eq. (\ref{tortoise coordinate}) to work with the Dirac equation in MSW background  

\begin{equation}\label{conformal metric}
ds^2=\xi d{\Tilde{s}}^2 = \xi (dt^2-dr_*^2)
\end{equation}

where $\xi$ is given in Eq. (\ref{xi}). Considering a conformal transformation, $g_{\mu\nu}=\xi \Tilde{g}_{\mu\nu}$, of the metric in Eq. (\ref{line element}), the relevant quantities in Dirac equation reduce to the following form \cite{Lopez-Ortega:2009flo}

\begin{equation}\label{transformation relation}
\Psi=\frac{1}{{\xi }^{1/4}}\Tilde{\Psi},\quad \quad \quad \slashed{\nabla}\Psi=\frac{1}{{\xi }^{3/4}}\Tilde{\slashed{\nabla}}\Tilde{\Psi}, \quad \quad \quad m_{eff}=\frac{1}{{\xi }^{1/2}}\Tilde{m}
\end{equation}

The Dirac equation (\ref{dirac equation cst}) reduces to a pair of coupled differential equations containing $\Tilde{\Psi}_1$ and $\Tilde{\Psi}_2$ which are components of the two dimensional Dirac spinor.
 
\begin{equation}\label{dirac equation}
\begin{split}
\partial_t\Tilde{\Psi}_2-\partial_{r_*}\Tilde{\Psi}_2=-im_{eff}\sqrt{\xi}\Tilde{\Psi}_1  \\
\partial_t\Tilde{\Psi}_1+\partial_{r_*}\Tilde{\Psi}_1=-im_{eff}\sqrt{\xi}\Tilde{\Psi}_2.
\end{split}
\end{equation}

Employing the same method as used for scalar perturbation we decompose the radial and temporal part of $\tilde{\Psi}$ as 
\begin{equation}\label{Psi in tort}
\Tilde{\Psi}_s(r_*,t)=R_s(r_*)e^{-i\omega t}, \quad \quad s=1,2
\end{equation}

Note that the function $\xi$ is dependent on $r_*$ only. As a result the temporal part exhibits a harmonic 
oscillator nature. 
Redefinition of $R_1$ by $e^{i\pi /4}\Tilde{R}_1$, and $R_2$ by $e^{-i\pi /4}\Tilde{R}_2$ reduces the above set 
of equations (\ref{dirac equation}) in the following form

\begin{equation}
\begin{split}
\frac{d\Tilde{R}_2}{dr_*}+i\omega \Tilde{R}_2=-m_{eff}\sqrt{\xi}\Tilde{R}_1 \\
\frac{d\Tilde{R}_1}{dr_*}-i\omega \Tilde{R}_1=-m_{eff}\sqrt{\xi}\Tilde{R}_2
\end{split}
\end{equation}

The QNM mode analysis is done by choosing linear combinations of $\Tilde{R}_1$ and $\Tilde{R}_2$ as $Z_{\pm}=\Tilde{R}_1\pm \Tilde{R}_2$. They will in turn give us the left and right chiral modes.  The equation for $Z_\pm$ can be written in a Schrodinger like form in  tortoise coordinate as

\begin{equation}
\frac{d^2Z\pm}{dr_*^2}+{\omega}^2Z_\pm =V_\pm(r_*) Z_\pm
\end{equation}

\begin{figure}[ht]
 \centering
   \includegraphics[width=6 cm, height =4 cm]{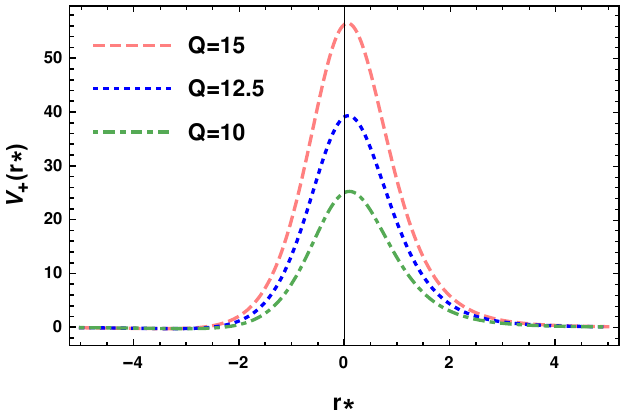}
  \caption{Effective potential experienced by $Z_+$ mode for different $Q$ in 
  tortoise coordinates with parameters $\sigma=-\frac{1}{2}$, $M=1$ and $\sqrt{k}=1$. }
 \label{fig:V(+)}
\end{figure}

\begin{figure}[ht]
 \centering
   \includegraphics[width=6 cm, height =4 cm]{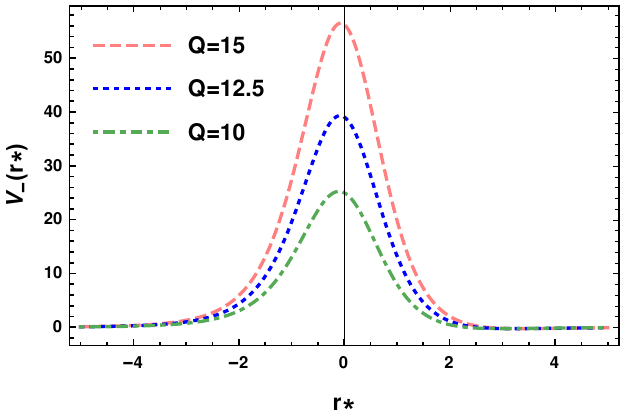}
  \caption{Effective potential experienced by $Z_-$ mode for different $Q$ in 
  tortoise coordinates with parameters $\sigma=-\frac{1}{2}$, $M=1$ and $\sqrt{k}=1$. }
 \label{fig:V(-)}
\end{figure}

Effective potentials  for the spinor modes   have the following form 
\begin{equation}\label{Dirac eff pot}
V_{\pm}(r_*)=Q \left(W(r_*)\right)^{\sigma -2} \left(\sqrt{k} Q e^{\frac{2 r_*}{\sqrt{k}}} \left(W(r_*)\right)^{\sigma +1}\mp\frac{\sqrt{e^{\frac{4 r_*}{\sqrt{k}}}} \left(2 \sqrt{k} \sigma  e^{\frac{2 r_*}{\sqrt{k}}}+M\right)}{\sqrt{1-\frac{M}{W(r_*)}}}\right)
\end{equation}

where $W(r_*)= \left(\sqrt{k} e^{\frac{2 r_*}{\sqrt{k}}}+M\right)$. The effective potentials experienced by the modes for different values of  constant $Q$ associated with the coupling term are plotted in Fig.   ({\ref{fig:V(+)}) and ({\ref{fig:V(-)}}}) with the choice of the parameters $\sigma=-\frac{1}{2}$, $M=1$ and $\sqrt{k}=1$. 
It is apparent from the figures that the effective potential will vanish at the horizon and at spatial infinity. The potentials also exhibit interesting features depending on different values $Q$. 

We now obtain decoupled equations for $R_1$ and $R_2$ from the coupled set of equations in Eq. (\ref{dirac equation}) 

\begin{equation}\label{eq: decoupled r}
\frac{1}{\xi}\frac{d^2 R_s}{dr_*^2 }+ \frac{m_{eff}}{\sqrt{\xi}}\frac{d}{dr_*} \left(\frac{1}{m_{eff}\sqrt{\xi}}\right)\frac{dR_s}{dr_*}+\frac{im_{eff}\omega \epsilon}{\sqrt{\xi}}\frac{d}{dr_*} \left(\frac{1}{m_{eff}\sqrt{\xi}}\right)R_s +\frac{{\omega}^2}{\xi}R_s=m^2_{eff}R_s
\end{equation}

where $\epsilon = \mp 1$ for $s = 1$ and $2$ respectively. Considering the coordinate transformation given in Eq. (\ref{x coordinate}) along with $\sqrt{k}=M$, $\sigma=-\frac{1}{2}$ and having  $\alpha_s$ and $\beta_s$ to satisfy the following relations

\begin{equation}
\alpha_s =  \left\{ \begin{array}{l} \frac{1}{2}\left( iM\omega \epsilon+1 \right)   \\ \\ -\frac{i}{2}M\omega \epsilon \end{array}\right. \qquad \qquad
\beta_s  =  \left\{ \begin{array}{l}  \frac{1}{2}\left(- iM\omega \epsilon+1 \right)  \\ \\  \frac{i}{2}M\omega \epsilon  \end{array}\right. 
\end{equation}

one can reduce Eq. (\ref{eq: decoupled r}) to  the standard hypergeometic differential equation  form by substituting $R_s(x)=x^{\alpha_s}(1-x)^{\beta_s}F_s(x)$

\begin{equation}\label{eq: Hypergeometric DE}
x(1-x)\frac{d^2F_s}{dx^2}+\left( c_s-\left( a_s+b_s+1 \right) x \right) \frac{dF_s}{dx}-a_sb_s F_s=0
\end{equation}

where
\begin{equation}
a_s=\alpha_s+\beta_s-\frac{iQ\sqrt{M}}{2}, \quad \quad \quad b_s=\alpha_s+\beta_s+\frac{iQ\sqrt{M}}{2}, \quad \quad \quad c_s=2\alpha_s+\frac{1}{2}.
\end{equation}
For $\epsilon=1$, the solution of Eq. (\ref{eq: Hypergeometric DE}) reveals the following form of $R_2$ 

\begin{equation}
\begin{split}
R_2 &= \tilde{C}_{1} x^{(iM\omega+1)/2} (1-x)^{iM\omega/2} {}_{2}F_{1} (a_2,b_2;c_2;x)  \\ 
& + \tilde{C}_{2}x^{-iM\omega /2}(1-x)^{iM\omega /2} {}_{2}F_{1} (a_2-c_2+1,b_2-c_2+1;2-c_2;x).
\end{split}
\end{equation}
We now apply the boundary condition that the modes are purely ingoing near the horizon. So we can choose $\tilde{C}_1=0$ to write $R_2$ in the following form 

\begin{equation}
R_2 = \tilde{C}_{2}x^{-iM\omega /2}(1-x)^{iM\omega /2} {}_{2}F_{1} (a_2-c_2+1,b_2-c_2+1;2-c_2;x)
\end{equation}
${}$\\
In the limit of spatial infinity $i.e.$ for $r_* \to + \infty$ ($x \to 1$)  we get $R_2$ in the following form by applying Kummer's property of the hypergeometric functions given in Eq. (\ref{kummer's property}).

\begin{equation} \label{eq: approximation radial infinity}
R_2  \approx  \frac{\Gamma(C_2)\Gamma(C_2-A_2-B_2)}{\Gamma(C_2-A_2)\Gamma(C_2-B_2)} e^{-i\omega r_*} + \frac{\Gamma(C_2)\Gamma(A_2+B_2-C_2)}{\Gamma(A_2)\Gamma(B_2)}e^{-\frac{r_*}{M}+i\omega r_* } 
\end{equation}

where $ A_2 = a_2-c_2+1,~B_2 = b_2-c_2+1,C_2 = 2-c_2.$ Imposing the condition that the modes are purely outgoing asymptotically, we obtain the following conditions 
\begin{equation}
C_2-A_2= -n \qquad \text{or} \qquad C_2-B_2 = -n, \qquad\text{where} \,\,n = 0, 1, 2, \dots 
\end{equation}

The  QNFs for the component $\Psi_2$ of the Dirac field turn out to be 

\begin{equation} \label{eq: QNF Dirac 2}
\omega_{I} = \frac{Q}{2\sqrt{M}}-\frac{i}{M}\left(n+\frac{1}{2}\right) \qquad \qquad
\omega_{II} = -\frac{Q}{2\sqrt{M}}-\frac{i}{M}\left(n+\frac{1}{2}\right)
\end{equation}

One can get the similar set of QNFs (\ref{eq: QNF Dirac 2}) for the component $\Psi_1$ of the Dirac field by applying the similar procedure. From these expressions of QNFs for both components $\Psi_1$ and $\Psi_2$, it is clearly seen that QNFs have both real and imaginary parts.z The real parts scale linearly with $Q$, constant  associated with coupling and imaginary parts are independent of that. On the other hand negative imaginary parts scale linearly with the overtone number, $n$ where as real parts are independent of $n$. QNFs up to overtone number $5$ have been presented in Table \ref{QNFS_s2} for different field couplings. 



\begin{figure}[ht]
\centering
\includegraphics[width=8.5 cm, height =4 cm]{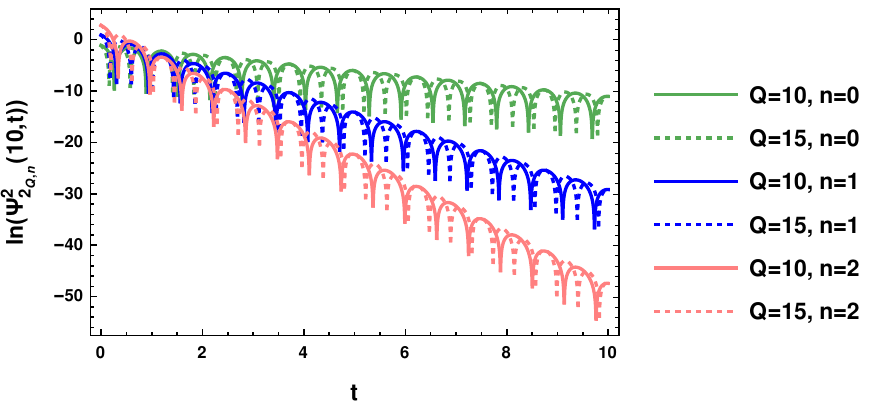}
\caption{In logarithmic scale the time evolution of square of $\Psi_2$ component of the Dirac field with different $Q$ values at $r=10$ corresponding to QNFs $\omega_I$ has been shown.}
\label{sqdqnm2log}
\end{figure}

\begin{figure}[ht]
\centering
\includegraphics[width=8.5 cm, height =4 cm]{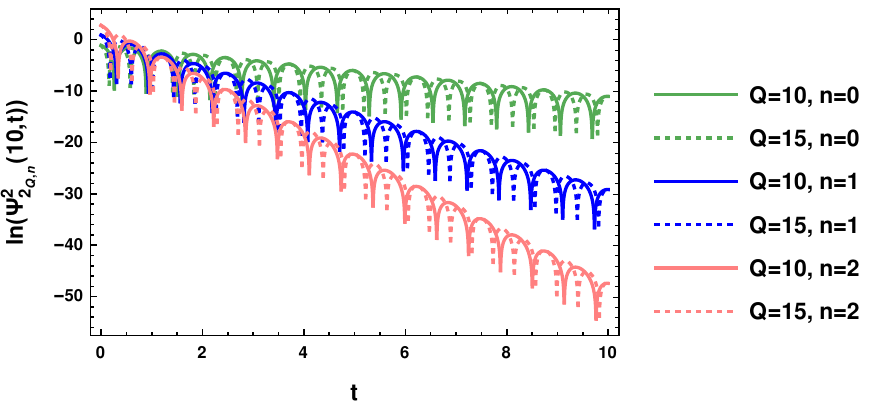}
\caption{Time evolution of square of $\Psi_2$ component of the Dirac field with different $Q$ values at $r=10$ corresponding to QNFs $\omega_{II}$ has been depicted in logarithmic scale. It is seen that as $Q$ increases the mode becomes more and more oscillatory for a fixed value of $n$. It is also clear that the damping of a particular mode increases with the increase of $n$.}
\label{sqdqnm2logwII}
\end{figure}

QNFs corresponding to Eq. (\ref{eq: QNF Dirac 2}) have both real and imaginary parts so the modes provide damped oscillatory behaviour during their temporal evolution at an observation position. Since the imaginary part of QNFs are independent of $Q$, the damping is also independent of the coupling strength.  On the other hand, the real part scales linearly with the coupling constant so depending on the strength of coupling the oscillation frequency of the modes will change. These features of the modes can be clearly seen in the Figure (\ref{sqdqnm2log}) and (\ref{sqdqnm2logwII}) for $\Psi_2$ component of Dirac field. Time evolution 
of modulus of $\Psi_2$  at a given position for different values of $Q$ is shown using a logarithmic scale. Figures depict that the damping of a mode increases with increasing overtone number. Modes from lager $Q$ value have larger oscillation frequency compared to other modes produced from lower $Q$ values for same overtone number, $n$.

\begin{table}[ht]
\caption{Real and imaginary parts of QNFs for the components ${{\Psi}}_{1}$ and ${{\Psi}}_{2}$ of the Dirac field for different values 
of $Q$ and $n$ considering $M=1$.}\label{QNFS_s2}
\begin{tabular*}{\textwidth}{@{\extracolsep\fill}lcccccc}
\toprule%
\toprule%
& \multicolumn{3}{@{}c@{}}{$Q=10$} & \multicolumn{3}{@{}c@{}}{$Q=15$} \\ \cmidrule{2-4} \cmidrule{5-7}%
$n$ & $Re\,\omega_I$ & $Re\,\omega_{II}$ & $Im\,\omega_I \, \text{or}\, Im\, \omega_{II}$ & $Re\,\omega_{I}$ & $Re\,\omega_{II}$ & $Im\,\omega_I \, \text{or}\, Im\, \omega_{II}$ \\
\midrule
$0$ &$ 5 $  & $-5 $ &$-0.5 $ &$7.5 $ &$-7.5 $ &$-0.5 $  \\
$1$ &$ 5 $  & $-5 $ &$-1.5 $ &$7.5 $ &$-7.5 $ &$-1.5 $  \\
$2$ &$ 5 $  & $-5 $ &$-2.5 $ &$7.5 $ &$-7.5 $ &$-2.5 $  \\
$3$ &$ 5 $  & $-5 $ &$-3.5 $ &$7.5 $ &$-7.5 $ &$-3.5 $  \\
$4$ &$ 5 $  & $-5 $ &$-4.5 $ &$7.5 $ &$-7.5 $ &$-4.5 $  \\
$5$ &$ 5 $  & $-5 $ &$-5.5 $ &$7.5 $ &$-7.5 $ &$-5.5 $  \\
\botrule
\toprule%
\end{tabular*}
\end{table}


\section{Conclusions}   
Blackhole quasi normal modes are of particular interest as they depend only on 
intrinsic blackhole parameters. They can be useful in estimating the parameters of the blackhole. The other importance 
lies in the fact that QNMs can be used in the semi-classical attempts to quantize a black hole area.
In this work there are two distinct results. The exact solutions of the quasi normal mode frequencies have been obtained for both 
Klein-Gordon and Dirac field. These can provide us with better understanding 
of the perturbation theory as they dictate the late time behaviour of the fields. 
In all the cases the modes scale with the blackhole parameters $k$ and $M$. They also depend on the mass of the perturbing Klein-Gordon field 
in general , however, the dependence can be negligibly small for largely damped modes under certain restrictions on the 
parameters of the theory.
For the Klein-Gordon field, we have chosen a non-minimal coupling of the dilaton 
field with the perturbing scalar to study the stability under scalar perturbations. It is worthy noting that the 
massless modes are supported by the non-minimal coupling term, {\textit{i.e.}} in the absence of this there will 
no such quasi normal mode in the scalar case. The QNMs are stable against these types of perturbations but 
there are possibilities for the existence of unstable modes depending on the masses of the perturbing 
fields and the coupling of the fields with the dilaton (for scalar perturbations). Here we have chosen a power law 
dependence, however, it may be interesting to explore the consequences of other kinds of coupling functions. 
It has been found for Klein-Gordon field, in a particular  
stable mode the most massive field experiences the least damping compared to other fields. {Damped oscillatory feature has been found in the time evolution profile of Dirac field and oscillation frequency increases with the of increase of coupling strength with the dilaton but damping is independent of it. For both types of fields, the damping feature increases with the increase of overtone number i.e. more the overtone number, a field is more damped.} 
\\

We propose to explore how these exact expressions of the quasi-normal frequencies 
can be used to find a relation between the highly damped modes and the quantum area of the black hole. As the 
QNFs are complex, in this case Maggiore's \cite{Maggiore:2007nq} proposal will be applicable.  The QNM spectrum in the heavily damped limit 
will be treated as a harmonic oscillator with frequency $\omega = \sqrt{\omega_{R}^2 + \omega_{I}^2}$. In future we would like 
to explore how QNMs can be related to the blackhole intrinsic microscopic properties. We further like to explore how 
the blackhole can be understood as an ensemble of harmonic oscillators exploiting the large $n$ limit of quasi normal modes.

\section{Acknowledgements}
P. Gayen acknowledges the University Grants Commission (UGC), India for providing financial support through a 
fellowship with ID: 191620137660.



\bibliographystyle{apsrev4-1}
\bibliography{references}

@article{Maggiore:2007nq,
    author = "Maggiore, Michele",
    title = "{The Physical interpretation of the spectrum of black hole quasinormal modes}",
    eprint = "0711.3145",
    archivePrefix = "arXiv",
    primaryClass = "gr-qc",
    doi = "10.1103/PhysRevLett.100.141301",
    journal = "Phys. Rev. Lett.",
    volume = "100",
    pages = "141301",
    year = "2008"
}

@article{Kokkotas:1999bd,
    author = "Kokkotas, Kostas D. and Schmidt, Bernd G.",
    title = "{Quasinormal modes of stars and black holes}",
    eprint = "gr-qc/9909058",
    archivePrefix = "arXiv",
    doi = "10.12942/lrr-1999-2",
    journal = "Living Rev. Rel.",
    volume = "2",
    pages = "2",
    year = "1999"
}

@article{Koley:2003tp,
    author = "Koley, Ratna and Pal, Supratik and Kar, Sayan",
    title = "{Geodesics and geodesic deviation in a two-dimensional black hole}",
    eprint = "gr-qc/0302065",
    archivePrefix = "arXiv",
    doi = "10.1119/1.1566426",
    journal = "Am. J. Phys.",
    volume = "71",
    pages = "1037--1042",
    year = "2003"
}

@book{abramowitz1964handbook,
  title={Handbook of mathematical functions},
  author={Abramowitz, Milton and Stegun, Irene A and others},
  address="New York",
  publisher="Dover",
  year={1964}
}

@article{Lopez-Ortega:2009flo,
    author = "Lopez-Ortega, A.",
    title = "{The Dirac equation in D-dimensional spherically symmetric spacetimes}",
    journal=" 	Lat. Am. J. Phys. Educ.",
    volume = "3",
    pages = "578",
    archivePrefix = "arXiv",
    primaryClass = "gr-qc",
    month = "6",
    year = "2009"
}

@book{wang1989special,
  title={Special functions},
  author={Wang, Zhu Xi and Guo, Dun Ren},
       address="Singapore",
    publisher = "World Scientific",
  year="1989"
}

@article{Verlinde:2016toy,
    author = "Verlinde, Erik P.",
    title = "{Emergent Gravity and the Dark Universe}",
    archivePrefix = "arXiv",
    primaryClass = "hep-th",
    doi = "10.21468/SciPostPhys.2.3.016",
    journal = "SciPost Phys.",
    volume = "2",
    number = "3",
    pages = "016",
    year = "2017"
}

@article{Tuveri:2019zor,
    author = "Tuveri, Matteo and Cadoni, Mariano",
    title = "{Galactic dynamics and long-range quantum gravity}",
    archivePrefix = "arXiv",
    primaryClass = "gr-qc",
    doi = "10.1103/PhysRevD.100.024029",
    journal = "Phys. Rev. D",
    volume = "100",
    number = "2",
    pages = "024029",
    year = "2019"
}

@article{Mandal:1991tz,
    author = "Mandal, Gautam and Sengupta, Anirvan M. and Wadia, Spenta R.",
    title = "{Classical solutions of two-dimensional string theory}",
    reportNumber = "IASSNS-HEP-91-10",
    doi = "10.1142/S0217732391001822",
    journal = "Mod. Phys. Lett. A",
    volume = "6",
    pages = "1685--1692",
    year = "1991"
}

@article{Witten:1991yr,
    author = "Witten, Edward",
    title = "{On string theory and black holes}",
    reportNumber = "IASSNS-HEP-91-12",
    doi = "10.1103/PhysRevD.44.314",
    journal = "Phys. Rev. D",
    volume = "44",
    pages = "314--324",
    year = "1991"
}

@article{Chan:1994qa,
    author = "Chan, K. C. K. and Mann, Robert B.",
    title = "{Static charged black holes in (2+1)-dimensional dilaton gravity}",
    archivePrefix = "arXiv",
    reportNumber = "WATPHYS-TH-94-01",
    doi = "10.1103/PhysRevD.50.6385",
    journal = "Phys. Rev. D",
    volume = "50",
    pages = "6385",
    year = "1994"
}

@article{Sebastian:2014dka,
    author = "Sebastian, Saneesh and Kuriakose, V. C.",
    title = "{Dirac Quasinormal modes of MSW black holes}",
    archivePrefix = "arXiv",
    primaryClass = "gr-qc",
    doi = "10.1142/S0217732314500199",
    journal = "Mod. Phys. Lett. A",
    volume = "29",
    number = "5",
    pages = "1450019",
    year = "2014"
}

@article{Fernando:2004ay,
    author = "Fernando, Sharmanthie",
    title = "{Greybody factors of charged dilaton black holes in 2 + 1 dimensions}",
    archivePrefix = "arXiv",
    reportNumber = "NKU-04-SF1",
    doi = "10.1007/s10714-005-0035-x",
    journal = "Gen. Rel. Grav.",
    volume = "37",
    pages = "461--481",
    year = "2005"
}

@article{Fernando:2003ai,
    author = "Fernando, Sharmanthie",
    title = "{Quasinormal modes of charged dilaton black holes in (2+1)-dimensions}",
    archivePrefix = "arXiv",
    reportNumber = "NKU-03-SF2",
    doi = "10.1023/B:GERG.0000006694.68399.c9",
    journal = "Gen. Rel. Grav.",
    volume = "36",
    pages = "71--82",
    year = "2004"
}

@article{Lopez-Ortega:2011ysu,
    author = "Lopez-Ortega, A. and Vega-Acevedo, I.",
    title = "{Quasinormal frequencies of asymptotically flat two-dimensional black holes}",
    archivePrefix = "arXiv",
    primaryClass = "gr-qc",
    doi = "10.1007/s10714-011-1185-7",
    journal = "Gen. Rel. Grav.",
    volume = "43",
    pages = "2631--2647",
    year = "2011"
}

@article{Zelnikov:2008rg,
    author = "Zelnikov, Andrei",
    title = "{Non-minimal scalar fields in 2D de Sitter and dilaton black holes}",
    archivePrefix = "arXiv",
    primaryClass = "hep-th",
    reportNumber = "ALBERTA-THY-13-08",
    doi = "10.1088/1126-6708/2008/07/010",
    journal = "JHEP",
    volume = "07",
    pages = "010",
    year = "2008"
}

@article{Cadoni:2021qfn,
    author = "Cadoni, Mariano and Oi, Mauro and Sanna, Andrea Pierfrancesco",
    title = "{Quasi-normal modes and microscopic description of 2D black holes}",
    archivePrefix = "arXiv",
    primaryClass = "gr-qc",
    doi = "10.1007/JHEP01(2022)087",
    journal = "JHEP",
    volume = "01",
    pages = "087",
    year = "2022"
}

@article{Grumiller:2002nm,
    author = "Grumiller, D. and Kummer, W. and Vassilevich, D. V.",
    title = "{Dilaton gravity in two-dimensions}",
    archivePrefix = "arXiv",
    reportNumber = "TUW-02-01",
    doi = "10.1016/S0370-1573(02)00267-3",
    journal = "Phys. Rept.",
    volume = "369",
    pages = "327--430",
    year = "2002"
}

@article{LISA:2022kgy,
    author = "Arun, K. G. and others",
    collaboration = "LISA",
    title = "{New horizons for fundamental physics with LISA}",
    archivePrefix = "arXiv",
    primaryClass = "gr-qc",
    doi = "10.1007/s41114-022-00036-9",
    journal = "Living Rev. Rel.",
    volume = "25",
    number = "1",
    pages = "4",
    year = "2022"
}

@article{Dvali:2020wqi,
    author = "Dvali, Gia",
    title = "{Entropy Bound and Unitarity of Scattering Amplitudes}",
    archivePrefix = "arXiv",
    primaryClass = "hep-th",
    doi = "10.1007/JHEP03(2021)126",
    journal = "JHEP",
    volume = "03",
    pages = "126",
    year = "2021"
}

@article{Dvali:2013eja,
    author = "Dvali, Gia and Gomez, Cesar",
    title = "{Quantum Compositeness of Gravity: Black Holes, AdS and Inflation}",
    archivePrefix = "arXiv",
    primaryClass = "hep-th",
    doi = "10.1088/1475-7516/2014/01/023",
    journal = "JCAP",
    volume = "01",
    pages = "023",
    year = "2014"
}

@article{Cadoni:2020mgb,
    author = "Cadoni, M. and Tuveri, M. and Sanna, A. P.",
    title = "{Long-Range Quantum Gravity}",
    archivePrefix = "arXiv",
    primaryClass = "gr-qc",
    doi = "10.3390/sym12091396",
    journal = "Symmetry",
    volume = "12",
    number = "9",
    pages = "1396",
    year = "2020"
}

@article{1985ApJ...291L..33S,
        author = "Schutz, B. F. and Will, C. M.",
        title = "{Black hole normal modes - A semianalytic approach}",
        doi = "10.1086/184453",
        journal = "Astrophys. J. Lett.",
        volume = "291",
        pages = "L33-L36",
        year = "1985"
}

@article{Chakraborty:2017qve,
    author = "Chakraborty, Sumanta and Chakravarti, Kabir and Bose, Sukanta and SenGupta, Soumitra",
    title = "{Signatures of extra dimensions in gravitational waves from black hole quasinormal modes}",
    archivePrefix = "arXiv",
    primaryClass = "gr-qc",
    doi = "10.1103/PhysRevD.97.104053",
    journal = "Phys. Rev. D",
    volume = "97",
    number = "10",
    pages = "104053",
    year = "2018"
}

@article{Kanti:2005xa,
    author = "Kanti, P. and Konoplya, R. A.",
    title = "{Quasi-normal modes of brane-localised standard model fields}",
    archivePrefix = "arXiv",
    doi = "10.1103/PhysRevD.73.044002",
    journal = "Phys. Rev. D",
    volume = "73",
    pages = "044002",
    year = "2006"
}

@article{Birmingham:2001pj,
    author = "Birmingham, Danny and Sachs, Ivo and Solodukhin, Sergey N.",
    title = "{Conformal field theory interpretation of black hole quasinormal modes}",
    archivePrefix = "arXiv",
    doi = "10.1103/PhysRevLett.88.151301",
    journal = "Phys. Rev. Lett.",
    volume = "88",
    pages = "151301",
    year = "2002"
}

@article{Kovtun:2005ev,
    author = "Kovtun, Pavel K. and Starinets, Andrei O.",
    title = "{Quasinormal modes and holography}",
    archivePrefix = "arXiv",
    reportNumber = "NSF-KITP-05-41",
    doi = "10.1103/PhysRevD.72.086009",
    journal = "Phys. Rev. D",
    volume = "72",
    pages = "086009",
    year = "2005"
}

@article{Seahra:2004fg,
    author = "Seahra, Sanjeev S. and Clarkson, Chris and Maartens, Roy",
    title = "{Detecting extra dimensions with gravity wave spectroscopy: the black string brane-world}",
    archivePrefix = "arXiv",
    doi = "10.1103/PhysRevLett.94.121302",
    journal = "Phys. Rev. Lett.",
    volume = "94",
    pages = "121302",
    year = "2005"
}

@article{Kettner:2004aw,
    author = "Kettner, Joanne and Kunstatter, Gabor and Medved, A. J. M.",
    title = "{Quasinormal modes for single horizon black holes in generic 2-d dilaton gravity}",
    eprint = "gr-qc/0408042",
    archivePrefix = "arXiv",
    doi = "10.1088/0264-9381/21/23/002",
    journal = "Class. Quant. Grav.",
    volume = "21",
    pages = "5317--5332",
    year = "2004"
}

@article{Bhattacharjee:2020nul,
    author = "Bhattacharjee, Srijit and Sarkar, Subhodeep and Bhattacharyya, Arpan",
    title = "{Scalar perturbations of black holes in Jackiw-Teitelboim gravity}",
    eprint = "2011.08179",
    archivePrefix = "arXiv",
    primaryClass = "gr-qc",
    doi = "10.1103/PhysRevD.103.024008",
    journal = "Phys. Rev. D",
    volume = "103",
    number = "2",
    pages = "024008",
    year = "2021"
}

@book{Hawking:1973uf,
    author = "Hawking, Stephen W. and Ellis, George F. R.",
    title = "{The Large Scale Structure of Space-Time}",
     address="New York",
    publisher = "Cambridge University Press",
    series = "Cambridge Monographs on Mathematical Physics",
    month = "2",
    year = "2023"
}

@inproceedings{Preskill:1992tc,
    author = "Preskill, John",
    title = "{Do black holes destroy information?}",
    booktitle = "{International Symposium on Black holes, Membranes, Wormholes and Superstrings}",
    archivePrefix = "arXiv",
    reportNumber = "CALT-68-1819",
    month = "1",
    year = "1992"
}

@article{MANN1990134,
author = "R.B. Mann and A. Shiekh and L. Tarasov",
title = "{Classical and quantum properties of two-dimensional black holes}",
journal = "Nucl. Phys. B",
volume = "341",
number = "1",
pages = "134-154",
year = "1990"
}

\end{document}